\documentclass[conference]{IEEEtran}
\IEEEoverridecommandlockouts
\usepackage{cite}
\usepackage{amsmath,amssymb,amsfonts}
\usepackage{algorithmic}
\usepackage{graphicx}
\usepackage{textcomp}
\usepackage[dvipsnames]{xcolor}
\usepackage{booktabs}
\def\BibTeX{{\rm B\kern-.05em{\sc i\kern-.025em b}\kern-.08em
    T\kern-.1667em\lower.7ex\hbox{E}\kern-.125emX}}
\usepackage{multirow,makecell}
\usepackage{pifont}

\newcommand{\verifycorrect}{[\textcolor{ForestGreen}{\ding{52}}]}
\newcommand{\verifyfail}{[\textcolor{Red}{\ding{56}}]}

\begin{document}
\normalsize

\title{RetroThinker: Enabling Retrospective Thinking in Speech LLMs}

\author{
    \IEEEauthorblockN{
        Yi-Jen Shih\IEEEauthorrefmark{1}, 
        Puyuan Peng\IEEEauthorrefmark{2},
        Abdelrahman Mohamed\IEEEauthorrefmark{2}, and 
        David Harwath\IEEEauthorrefmark{1}
    }
    \IEEEauthorblockA{\IEEEauthorrefmark{1}\textit{The University of Texas at Austin}}
    \IEEEauthorblockA{\IEEEauthorrefmark{2}\textit{FAIR, Meta Superintelligence Labs}}
    \thanks{All data access, experiments, and processing activities were conducted by The University of Texas at Austin.}
}

\maketitle

\begin{abstract}
Speech large language models (SpeechLLMs) offer reduced latency and retain paralinguistic nuances that are typically lost in cascaded automatic speech recognition (ASR) and text-based LM architectures. However, they continue to lag behind text-only LLMs on complex reasoning tasks, while real-time spoken interaction imposes strict latency constraints. Although prior works employ Chain-of-Thought (CoT) and concurrent reasoning to enhance reasoning capabilities without inducing prohibitive delays, an inherent accuracy-latency trade-off persists. In this paper, we investigate whether a streaming SpeechLLM can dynamically revise its reasoning traces on the fly. We introduce \textit{RetroThinker}, a multi-stage post-training framework that equips the Moshi model to self-verify and forward-correct CoT steps during inference. RetroThinker combines supervised fine-tuning (SFT) on curated retrospective thinking data with length-based direct preference optimization (DPO) to optimize retrospective during early reasoning (i.e., reasoning concurrently while the user speaks). Evaluated on the GSM8K benchmark, RetroThinker significantly improves the accuracy-latency trade-off over non-retrospective baselines, achieving an 11\% absolute accuracy gain at a comparable latency.
\end{abstract}

\begin{IEEEkeywords}
Speech LLMs, Reasoning
\end{IEEEkeywords}

\section{Introduction}
In recent years, AI agents have become pervasive, offering diverse capabilities seamlessly integrated into personal devices like smartphones and smart glasses.
Among all modalities, speech serves as the most direct and natural interface for human-agent interaction.
Unlike text, speech conveys substantially more information, capturing speaker identity, prosody, and emotional nuances typically lost in transcription.

Most contemporary AI agents rely on Large Language Models (LLMs) as their backbone and are trained on massive text corpora.
Consequently, enabling them to process speech generally follows one of the two paradigms.
The first is a cascaded approach~\cite{chipchat,arora25_espsds,chen2025firered}, which concatenates modular components: Automatic Speech Recognition (ASR), a text-based LLM, and Text-to-Speech synthesis (TTS).
The second is to pretrain and fine-tune LLMs to directly consume and generate speech tokens.
This approach, which we term SpeechLLMs~\cite{kyutai24_moshi, zeng24_glm4voice, roy26_personaplex}, bypasses the modular constraints and information bottlenecks inherent in cascaded systems.
The primary advantages of SpeechLLMs include lower inference latency, simplified deployment, and a superior capacity to process paralinguistic signals.
As a result, they are uniquely equipped to facilitate naturalistic conversations between humans and machines.

Despite the inherent advantages of SpeechLLMs, two significant challenges remain. First, recent studies reveal that they often underperform in reasoning and general intelligence when compared to text-based counterparts of similar scale and training data volume~\cite{shih26_canspeechllm,xiang_SpeechLLM_modality_gap}.
This performance disparity is particularly pronounced in complex reasoning tasks, as opposed to simple factual retrieval.
The second challenge is that response latency is far more constrained than in text-based LLMs, as users expect near real-time interactions with voice agents.

Motivated by Chain-of-Thought (CoT) prompting~\cite{wei22_cot}, recent works~\cite{shih26_canspeechllm, chiang26_SHANKS,chiang26_STITCH,Xie25_omni_reason} integrate CoT into SpeechLLMs to enhance their conversational reasoning capabilities.
Originally introduced in the NLP community, CoT is widely adopted to boost model performance by eliciting step-by-step reasoning.
By decomposing complex problems into intermediate logical steps, CoT significantly improves accuracy across diverse benchmarks.
However, when applied to SpeechLLMs, generating the additional CoT trace substantially increases latency.

\begin{figure}[t]
  \centering
\includegraphics[width=\linewidth,trim={0.9cm 0.42cm 0.976cm 0.03cm},clip]{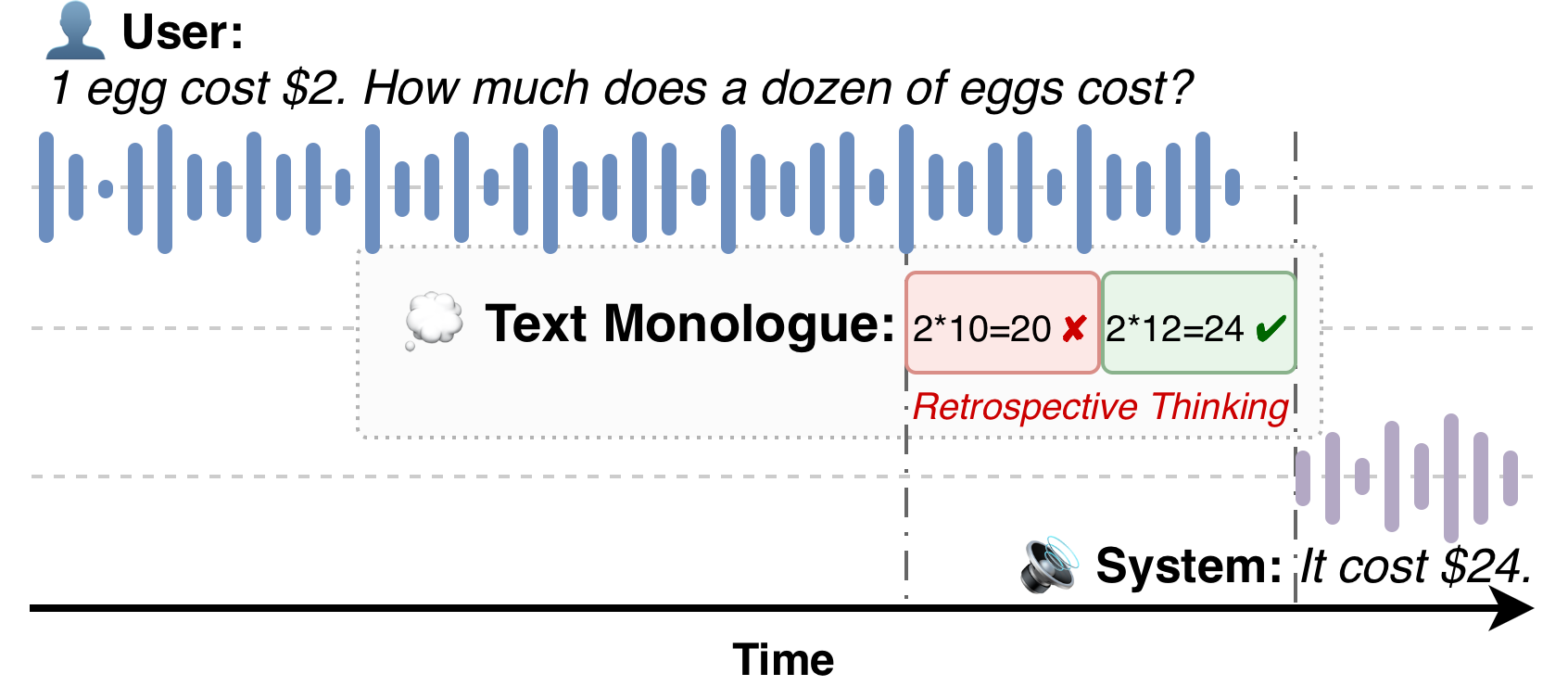}
  \caption{Overview of RetroThinker. By combining EarlyReasoning~(reasoning starts before the user finishes the question) and Retrospective Thinking~(self-verification and forward correction), the model achieves a better accuracy-latency trade-off.}
  \label{fig:overview}
  \vspace{-20pt}
\end{figure}

To mitigate this, prior work reduces latency by enabling concurrent reasoning: models either initiate the reasoning process while the user is still speaking~\cite{shih26_canspeechllm,chiang26_SHANKS}, or simultaneously reason while generating speech output~\cite{chiang26_STITCH,Xie25_omni_reason}.
This paradigm shift allows SpeechLLMs to process audio concurrently with reasoning.
While this improves reasoning and reduces response latency, an inherent accuracy-latency Pareto trade-off remains~\cite{lin2025voiceevaluationreasoningability}.

To push this accuracy-latency Pareto frontier further, we explore a parallel paradigm: \textit{retrospective thinking}.
In human conversation, when addressing complex inquiries, individuals occasionally commit logical errors during their internal reasoning processes.
However, humans possess the cognitive flexibility to reflect on these thoughts and perform real-time revisions during the dialogue~\cite{HOFFMAN1983_retro}.
Motivated by this behavior, we introduce the concept of retrospective thinking to SpeechLLMs.
Much like their human counterparts, it is inevitable that SpeechLLMs will introduce errors within their reasoning traces during inference, particularly under strict real-time constraints.
Consequently, we posit that retrospective thinking can improve reasoning accuracy by allowing the model to identify and rectify such errors dynamically.

We propose \textbf{RetroThinker}, a post-training framework that equips SpeechLLMs with real-time retrospective thinking capabilities.
Adapting retrospective sampling~\cite{wu25_retro}, we apply specialized fine-tuning techniques to SpeechLLMs, enabling them to identify incorrect CoT steps and subsequently revise them.
Following~\cite{shih26_canspeechllm}, we select Moshi~\cite{kyutai24_moshi} as our foundational SpeechLLM.
Its native multistream architecture offers the flexibility to inject CoT reasoning into the text stream without disrupting the synchronized audio input and output streams.
Furthermore, to mitigate the additional latency introduced by the retrospective thinking process, we investigate its interplay with two latency-reduction techniques from prior work~\cite{shih26_canspeechllm}: Early Reasoning (i.e., ``thinking while listening'') and Direct Preference Optimization (DPO)~\cite{rafailov2023_dpo}.
Evaluations on TTS-spoken GSM8K demonstrate that combining all three techniques improves the accuracy-latency trade-off in a controlled multi-step reasoning setting.
Additionally, we conduct an in-depth analysis to elucidate how these techniques influence one another.
In summary, our key contributions are as follows:
\begin{enumerate}
    \item We introduce retrospective thinking for streaming SpeechLLMs, enabling forward-only verification and correction of hidden CoT traces without rolling back generated speech.
    \item We provide a concrete post-training recipe for Moshi that combines rule-based initialization, sample-based alignment to model errors, and length-based DPO.
    \item On TTS-spoken GSM8K, RetroThinker improves the accuracy-latency frontier, outperforming a comparable non-retrospective baseline by 11 percentage points in accuracy at comparable latency.
\end{enumerate}

\section{Background}
\subsection{Moshi}
Moshi \cite{kyutai24_moshi} is a multi-stream, full-duplex SpeechLLM.
During each forward pass, the model concurrently processes three distinct streams: User Audio, System Text (a.k.a., inner monologue), and System Audio. 
To process these heterogeneous inputs as discrete sequences, Moshi leverages the Mimi codec to discretize continuous speech waveforms into audio tokens at 80 ms intervals using 8 codebooks.
The inner monologue stream provides word-level, time-aligned transcripts synchronized with the system’s speech output. 
Architecturally, Moshi utilizes a Temporal Transformer coupled with a Depth Transformer. 
At each timestep, the Temporal Transformer consumes all three input streams to predict the next token in the system text stream.
Subsequently, the Depth Transformer takes the Temporal Transformer’s output to predict the 8 audio tokens for the system audio stream. 
The temporal transformer undergoes initial text-only pre-training before being integrated into Moshi. 
Finally, the complete model is pre-trained on large-scale audio datasets and fine-tuned on multi-stream spoken conversational data.

\subsection{CoT Finetuning and Early Reasoning}
In \cite{shih26_canspeechllm}, the authors enhance Moshi's reasoning capabilities while reducing latency through early reasoning.
They achieve this by first relaxing the strict alignment between the system's audio and text streams. 
Specifically, the model is trained to generate a CoT trace in the text stream after user audio question and before producing the spoken response.
To reduce the latency added by CoT generation, they introduce a ``thinking-while-listening'' framework that initiates reasoning before the user finishes speaking.
To determine the optimal moment to start reasoning, they propose the \textit{Question Completeness (QC)} metric, which evaluates partial question lengths. QC ranges from 0 (empty) to 1 (complete). 
By defining a threshold $\theta \in \left[0,1\right]$, training data is curated to trigger reasoning once sufficient information is gathered. 
On inference, the model naturally initiates early reasoning by emitting a special token, functioning independently of the QC metric. 
Thus, tuning $\theta$ enables explicit control over the accuracy-latency trade-off. 
Furthermore, they proposed \textit{LengthDPO} to reduce latency under the Early Reasoning scenario. Specifically, they construct preference pairs where the positive samples are both accurate and concise, while the negative samples are incorrect, thereby minimizing latency without compromising performance. In our work, we define \textit{Standard Reasoning} as the setting where the CoT trace is generated only after the full user query is completed. 
Conversely, we define \textit{Early Reasoning} as the setting where the model begins its reasoning process concurrently while the user is still speaking.

\subsection{Retrospective Sampling}
In REVERSE~\cite{wu25_retro}, the authors mitigate hallucinations in Vision-Language Models~(VLMs) on image captioning task by training them with retrospective sampling abilities.
Within their framework, captions are first partitioned into multiple segments. 
Subsequently, a combination of rule-based methods and LLM prompting is employed to synthesize error caption segments.
These error segments are then randomly inserted before the corresponding ground-truth segments. 
At the end of each segment, a specialized self-validation token is appended to indicate the accuracy of the preceding segment.
During the training phase, the next-token prediction loss is masked for the injected error segments to prevent the model from learning incorrect generation patterns. 
During inference, the authors introduce a mechanism termed ``Retrospective Sampling.'' 
If the self-verification token identifies a segment as incorrect, the system rolls back the generation to a specific point and employs various strategies to regenerate the segment, ensuring factual consistency.

\begin{figure*}[t]
  \centering
\includegraphics[width=\linewidth,trim={0.7cm 0.2cm 0.7cm 0.2cm},clip]{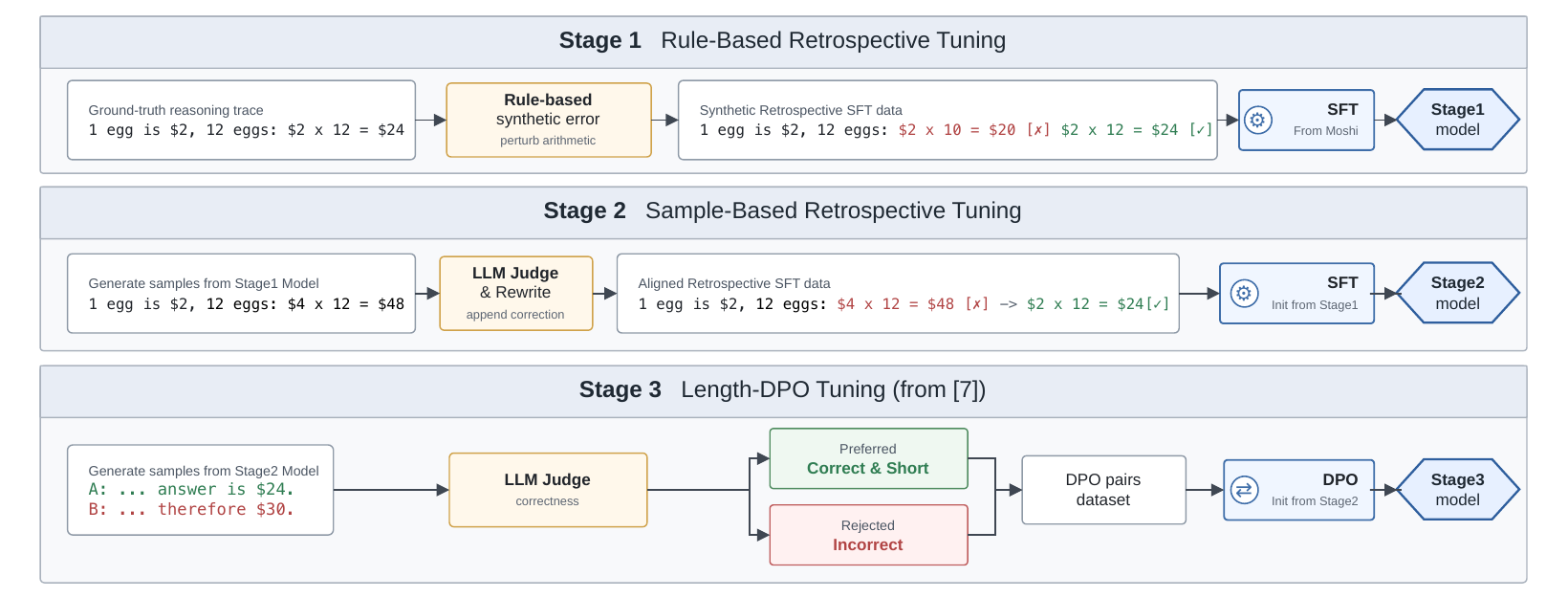}
  \caption{Illustration of the three-stage RetroThinker pipeline. Stage 1 employs a rule-based method to inject synthetic errors, teaching the model proper self-validation and self-correction formats. Stage 2 utilizes an external LLM judge to annotate the authentic errors generated by sampling the Stage 1 model. Finally, in Stage 3, LengthDPO is applied to mitigate the distribution mismatch caused by the external judge and to condense unnecessarily long retrospective reasoning traces.}
  \label{fig:retrothinker_pipeline}
\end{figure*}

\section{RetroThinker}
\label{sec:retrothinker}
We propose \textbf{RetroThinker}, a multi-stage framework~(See Fig.~\ref{fig:retrothinker_pipeline}) to enable retrospective thinking in SpeechLLMs.
Built upon the architecture in~\cite{shih26_canspeechllm}, our framework modifies the reasoning trace while maintaining the integrity of the remaining streams.
Consequently, our method is compatible with both \textit{standard} and \textit{early reasoning}, and can be combined with other concurrent-reasoning methods~\cite{chiang26_STITCH,chiang26_SHANKS,Xie25_omni_reason}.

\noindent
\textbf{Stage 1: Rule-based Retrospective Training.} We first segment each GSM8K ground-truth rationale into ordered CoT steps using sentence and equation boundaries, keeping the final answer as the spoken response target.
For each selected step, we synthesize an erroneous variant by applying rule-based perturbations, such as modifying numerical values or substituting arithmetic operators.
The training sequence then contains the erroneous step, a negative self-verification marker, the corrected step, and a positive marker before continuing to the next step.
The verification markers are rendered as \verifyfail{} and \verifycorrect{} in our qualitative examples.
Unlike REVERSE~\cite{wu25_retro}, which performs rollback or terminates the sequence after an incorrect segment, our framework requires the model to continue the generation process.
In streaming speech applications, the system cannot perform a temporal ``rollback'' without disrupting the user experience.
Therefore, we structure our data such that the model learns to identify, revise, and proceed with subsequent CoT steps even when interleaved with previous errors.
We perform Supervised Fine-Tuning (SFT) on Moshi using this data, masking the loss on the synthesized erroneous CoT tokens while retaining loss on the verification markers and corrected continuation, similarly to~\cite{wu25_retro}.

\noindent
\textbf{Stage 2: Sample-based Distribution Alignment.} While Stage 1 establishes the structural format for retrospective thinking, rule-based errors often fail to reflect the actual distribution of model failures during inference.
To mitigate this distribution mismatch, we proposed a second stage of sample-based fine-tuning.
We generate reasoning traces from the Stage 1 checkpoint and use an LLM-as-a-judge to evaluate each step given the question and preceding reasoning context.
Whenever the judge identifies an incorrect step, it returns a corrected continuation from that point to the final answer.
We then construct retrospective training examples by using the model-generated error as the masked negative step and the continuation provided by the judge as the positive correction.
This keeps the same loss-masking strategy as Stage 1 while aligning the correction behavior to authentic model failures.

\noindent
\textbf{Stage 3: Length-based DPO Fine-tuning.} While retrospective thinking significantly enhances accuracy, the inclusion of revision steps non-trivially increases the total thinking length, thereby escalating response latency.
To address this, we employ LengthDPO~\cite{shih26_canspeechllm} to compress the reasoning trace without sacrificing accuracy.
Specifically, we sample outputs from the Stage 2 model and construct preference pairs where preferred examples are correct and concise, while dispreferred examples are incorrect.
Beyond latency reduction, DPO fine-tuning serves as a critical mechanism for reducing the distribution mismatch inherent in \textit{early reasoning}~\cite{shih26_canspeechllm}.
In Stage 2, the reasoning traces are judged and revised by an LLM without exposure to the constraint that the user's query may not yet be fully received.
By applying DPO in the \textit{early reasoning} setting, the model learns to adapt its retrospective capabilities to the uncertainties of incomplete user queries.

\noindent
\textbf{Inference Strategy.} Departing from the rollback mechanism in~\cite{wu25_retro}, our inference procedure adheres to the streaming constraints of SpeechLLMs.
Our model performs ``forward-only'' self-correction: it continues generating subsequent CoT steps and the final response while preserving the entire reasoning history, including both incorrect and corrected steps, in the text stream.
If the model emits a negative verification marker, correction is produced as ordinary subsequent text.
This setup matches the way we finetune the model.

\section{Experiments}
\subsection{Training and Inference Details}
\noindent
\textbf{Training Details.}
To minimize computational cost, we employ LoRA~\cite{hu2022lora} rather than full fine-tuning~\cite{shih26_canspeechllm} of the Moshi model.
We select GSM8K~\cite{cobbe21_gsm8k} as our controlled training and evaluation dataset because it provides high-quality multi-step ground-truth reasoning steps and objective final answers.
These properties makes it suitable for us to evaluate the effectiveness of our approaches.
To adapt this text-based dataset for the speech domain, we synthesize the questions and answers into audio using Gemini-Flash-TTS.
We emphasize that this setup evaluates spoken delivery of multi-step math reasoning rather than natural open-domain dialogue; broader speech benchmarks such as VoiceBench~\cite{chen2024voicebench} contain reasoning examples but mix them with many factual or short-form tasks that do not require retrospective CoT correction.
For all SFT stages~(Stage 1 and Stage 2), we search learning rates in the range \texttt{2e-5} to \texttt{2e-6}, selecting checkpoints based on the minimum validation loss.
For the DPO phase, we search learning rates between \texttt{1e-7} and \texttt{1e-6} with $\beta=0.1$.
Additionally, to stabilize preference learning, we add an additional Negative Log-Likelihood (NLL) loss (weighted by $0.1$) on the positive examples~\cite{Xu_cpo}, normalize the probability distribution by length~\cite{meng2024_simpo}, and use only the text stream to calculate the policy distribution~\cite{Wu2025AligningSD}.
The final checkpoints are selected based on the highest reward-based accuracy on the validation set.

\noindent
\textbf{Evaluation Details.}
We adopt the same evaluation pipeline as~\cite{shih26_canspeechllm}.
During inference, we stream the user's audio query as input into Moshi and capture the model's response waveform.
All reported settings share the same synthesis, decoding, transcription, and judging pipeline.

\noindent
\textbf{Latency Metrics.}
To quantify response latency, we apply Voice Activity Detection (VAD) to the output waveform. Latency is defined as the number of seconds between the end of the user's speech and the onset of Moshi's audio response.

\noindent
\textbf{Accuracy Calculation.}
Following~\cite{shih26_canspeechllm}, we transcribe the generated audio using Whisper~\cite{Radford23_whisper} for accuracy evaluation.
We evaluate the generated waveform rather than Moshi's internal text stream because the waveform is the user-facing output.
The resulting transcripts are then evaluated using an LLM-as-a-Judge framework to determine logical correctness.
Across all experiments, we utilize \texttt{Qwen3-235B-Instruct}~\cite{qwen3technicalreport} as our LLM.

\subsection{Baselines and Experiment Settings}
To investigate the interplay between RetroThinker, \textit{early reasoning}, and \textit{LengthDPO}, we establish a comprehensive experimental matrix across two primary scenarios: \textit{Standard Reasoning} and \textit{Early Reasoning}.
For early reasoning, we run experiments with 4 different QC thresholds: $\theta\in \{65\%, 75\%, 85\%, 95\%\}$.
For both scenarios, we conduct comparative experiments with and without RetroThinker.
Finally, we apply LengthDPO to the early-reasoning models to examine the effect of reducing the reasoning trace on overall system latency.
We regard all settings without RetroThinker as our baselines.

\begin{figure}[t]
  \centering
  \includegraphics[width=\linewidth,trim={0.2cm 0.2cm 0.2cm 0},clip]{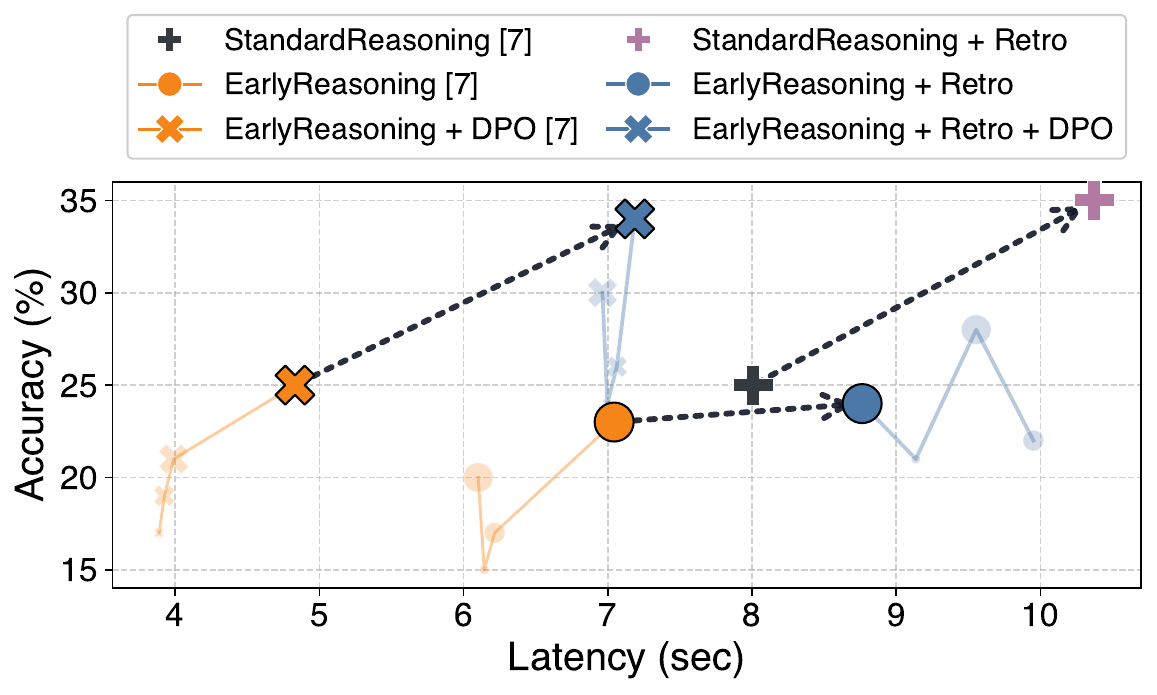}
  \caption{Accuracy-latency trade-off for different configurations. For Early Reasoning settings, varying Question Completeness (QC) thresholds $\theta \in \{65\%, 75\%, 85\%, 95\%\}$ are represented by different marker sizes; an increase in size denotes a higher threshold. The 95\% threshold is highlighted. Gray dotted arrows indicate the performance shift from adding RetroThinker. The baselines are he settings without Retro.}
  \label{fig:acc_lat_plot}
\end{figure}

\subsection{Main Results}
The accuracy and latency performance for each setting is illustrated in Figure~\ref{fig:acc_lat_plot}.
Our analysis yields several key insights.

\noindent
\textbf{RetroThinker generally improves accuracy at the cost of latency.}
As indicated by the gray arrows, integrating RetroThinker consistently improves accuracy across the evaluated settings. However, these improvements come at the cost of increased latency before applying LengthDPO, averaging approximately two seconds.

\noindent
\textbf{RetroThinker improves accuracy for early reasoning with the help of DPO.}
Furthermore, the observed improvement in \textit{EarlyReasoning} is limited when compared to \textit{StandardReasoning}.
This aligns with our hypothesis in Section~\ref{sec:retrothinker} (Stage 3), which posits a mismatch when using an external LLM to evaluate and rewrite model-generated samples.
By applying DPO, we largely alleviate this discrepancy, yielding accuracy and latency trade-offs comparable to standard reasoning.

\noindent
\textbf{Combining RetroThinker with EarlyReasoning and LengthDPO gives the best trade-off.}
Our best setting, \textit{EarlyReasoning} + Retro + DPO with the 95\% QC threshold, achieves an 11 percentage-point absolute accuracy improvement while keeping nearly the same latency compared to Standard Reasoning.
It also obtains about a 30\% latency reduction compared to \textit{StandardReasoning} + Retro with comparable accuracy.
In contrast, removing the RetroThinker component (leaving only \textit{EarlyReasoning + DPO}, as in \cite{shih26_canspeechllm}) results in a substantial accuracy drop.
These results indicate that RetroThinker is a critical driver of accuracy, and its combination with \textit{Early Reasoning} and DPO yields the most favorable accuracy-latency Pareto frontier.

\noindent
\textbf{Latency control of the QC metric is diminished with RetroThinker.}
Despite the improved Pareto frontier achieved with RetroThinker, we observe that the QC metric loses its precise control over latency compared to models without retrospective thinking.
In analyzing this phenomenon, we identify two primary factors affecting the final latency: how early the model initiates the reasoning trace and the total length of the reasoning trace.
We find that while the onset of reasoning is strictly dictated by the QC threshold (i.e., a higher threshold delays initiation), the resulting CoT lengths remain highly variable.
We leave explicit control of CoT lengths under retrospective thinking to future work.

\section{Detailed Analysis}
\begin{table}[t]
\centering
  \caption{Examples of different self-correction behaviors. The bold question text is the content spoken by the user before EarlyReasoning begins. The symbols~\verifycorrect~and~\verifyfail~indicate the model's self-verification result.}
  \label{tab:qualitative}
  \centering
  \begin{tabular}{ r p{0.35\textwidth} }
    \toprule
      \multicolumn{2}{l}{\textbf{Revision is based on new information after the start of reasoning}}  \\
    \midrule
    Question: & \textbf{The rug is 5 feet wider than the chair.} The couch is 2 feet longer than twice the width of the rug. If the chair is 3 feet wide. How many feet long is the couch? \\
    Reasoning: &  The rug is 5 feet wide~\verifyfail~The rug is 5 feet wider than the chair, so it is 3 + 5 = 8 feet wide.~\verifycorrect~ The couch is 2 feet longer than twice the width of the rug, so it is 2 * 8 + 2 = 18 feet long.~\verifycorrect \\
    \midrule
     \multicolumn{2}{l}{\textbf{Revision is purely based on prior step's logical mistake}}  \\
    \midrule
    Question: & \textbf{Two girls each got 1/6 of the 24 liters of water. Then a boy got 6 liters of water. How many }liters of water were left? \\
    Reasoning: &

The two girls got 1/6 * 24 = 4 liters of water.~\verifyfail~The two girls each got 1/6 * 24 = 4 liters of water.
So, they got 4 + 4 = 8 liters of water.~\verifycorrect~The boy got 6 liters, so the total amount of water taken is 8 + 6 = 14 liters.~\verifycorrect~The remaining water is 24 - 14 = 10 liters.
     \\
    \bottomrule
  \end{tabular}
\end{table}

\begin{table}[t]
  \caption{Ablation results for Stage 1 (rule-based) and Stage 2 (sample-based) SFT across different QC thresholds ($\theta$).}
  \label{tab:rule_sample_retro}
  \centering
  \begin{tabular}{ r c c c c c c c c}
    \toprule
    \textbf{Settings}& \multicolumn{4}{c}{\textbf{Accuracy (\%)}} & \multicolumn{4}{c}{\textbf{Latency (seconds)}}  \\
    \cmidrule(l{8pt}r{8pt}){2-5} \cmidrule(l{8pt}r{8pt}){6-9}
    $\theta$(\%) & 65 & 75 & 85 & 95 & 65 & 75 & 85 & 95 \\
    \midrule
    EarlyReasoning & 15 & 17 & 20 & 23 & 6.1 & 6.2 & 6.1 & 7.0 \\
    + Stage~1 & 13 & 16 & 19 & 23 & 7.7 & 8.0 & 6.6 & 8.4 \\
    + Stage~1\&2 & 21 & 22 & 28 & 24 & 9.1 & 10.0 & 9.6 & 8.8 \\
    \bottomrule
  \end{tabular}
\end{table}

\begin{table}[t]
\setlength{\tabcolsep}{5pt}
  \caption{CoT lengths and RetroThinking Ratio before and after LengthDPO for the EarlyReasoning + Retro setting.}
  \label{tab:thnk_len_retro_percent}
  \centering
  \begin{tabular}{ r cccc cccc}
    \toprule
    \textbf{Settings}& \multicolumn{4}{c}{\textbf{CoT Length (\# of tokens)}} & \multicolumn{4}{c}{\textbf{RetroThinking Ratio (\%)}}  \\
    \cmidrule(l{8pt}r{8pt}){2-5} \cmidrule(l{8pt}r{8pt}){6-9}
    $\theta$(\%) & 65 & 75 & 85 & 95 & 65 & 75 & 85 & 95 \\
    \midrule
    Before DPO  & 158.6 & 157.7 & 138.9 & 133.9 & 75.7 & 80.3 & 69.5 & 65.1 \\
    After DPO   & 134.8 & 125.5 & 111.2 & 99.8  & 59.5 & 58.2 & 30.5 & 7.2 \\
    \bottomrule
  \end{tabular}

\end{table}

\subsection{Understanding the Accuracy Boost of RetroThinker}
To understand how retrospective thinking dynamically improves accuracy, we categorize model revisions into two types: (1) correcting previous logical errors, and (2) integrating new information from the ongoing user query. Utilizing an LLM for automated classification on the \textit{EarlyReasoning} ($\theta=65\%$) + Retro + DPO setup, we find that 23\% of revisions are Type 1 and 32\% are Type 2. This suggests the model often refines its reasoning based on incoming user context.
We provide one example for each type in Table~\ref{tab:qualitative}.
However, in 42\% of instances, the revision remains logically identical to the original step, revealing that incorrect verification still causes redundant retrospective thinking. Addressing this inefficiency remains a clear direction for future work.

\subsection{Rule-based vs. Sample-based RetroThinker}
RetroThinker conducts SFT on retrospective thinking in two stages (Section~\ref{sec:retrothinker}).
As shown in Table~\ref{tab:rule_sample_retro}, accuracy remains stagnant after Stage 1 (rule-based).
This indicates that while Stage 1 establishes the structural format of retrospective tokens, it fails to impart strong self-correction abilities.
In contrast, Stage 2 (sample-based) leverages authentic model-generated errors, yielding significant accuracy improvements—notably a 9\% absolute gain at the $\theta=0.85$ threshold.
This highlights the necessity of sample-based training to bridge the gap between synthetic perturbations and actual inference errors.

\subsection{Identifying the Accuracy Bottleneck}
Effective retrospective thinking requires both self-verification of CoT steps and subsequent self-correction.
To disentangle these, we replace the model's internal verification tokens with an external LLM as an ``oracle verifier.''
During inference, whenever a verification token is sampled, the oracle evaluates the preceding reasoning and replaces the model's token with its judgment.
To establish an upper bound, we apply this to our most accurate setting (\textit{Standard Reasoning + Retro}).
The oracle improves accuracy from 35\% to 42\%. This 7\% gain demonstrates that while some errors stem from internal verification failures, the remaining 58\% error rate indicates the primary bottleneck is the error revision process itself.
We leave the refinement for future research.

\subsection{Impact of LengthDPO on Retrospective Thinking}
Beyond mitigating distribution mismatch (Section~\ref{sec:retrothinker}), we analyze how LengthDPO alters generation by examining CoT length and the Retrospective Thinking Ratio (the proportion of test instances with at least one CoT revision).

As shown in Table~\ref{tab:thnk_len_retro_percent}, average CoT length and the Retrospective Thinking Ratio decrease consistently across all QC thresholds.
The ratio's reduction is more pronounced at higher $\theta$.
Intuitively, limited initial information (lower $\theta$) causes more reasoning errors, whereas more complete input (higher $\theta$) enables robust initial hypotheses, requiring fewer revisions.
Coupled with the observed accuracy gains, these findings suggest that DPO finetuning makes the model more selective.
By retaining only essential retrospective steps, this optimization reduces response latency without sacrificing accuracy.

\subsection{Analysis by Reasoning Complexity}
\begin{figure}[t]
      \centering
      \includegraphics[width=\linewidth]{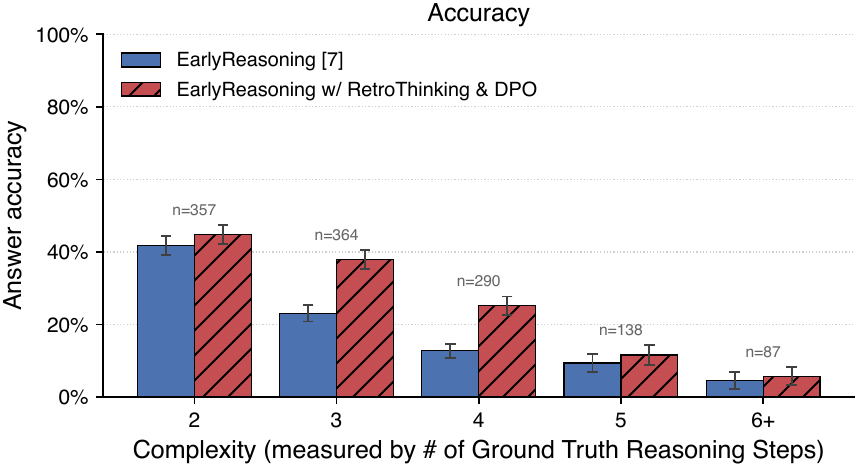}\\[0.5em]
      \includegraphics[width=\linewidth]{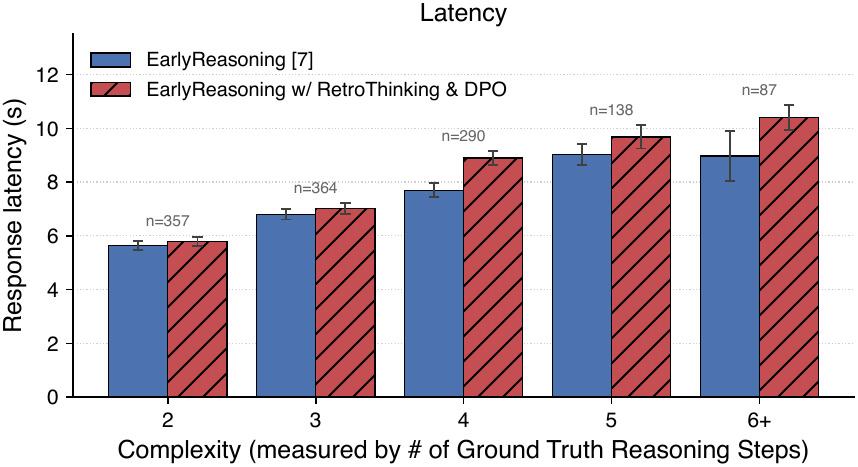}
      \caption{Accuracy (top) and latency (bottom) grouped by the question complexity~(number of
      ground-truth reasoning steps). Error bars denote $\pm1$ standard error
      of the mean.}
      \label{fig:acc-lat-by-steps}
  \end{figure}

To understand how performance scales with problem difficulty, we group the
GSM8K test questions by its complexity, measured
by the number of intermediate steps in the reference solution.
We focus on multi-step problems ($2$--$6{+}$ steps) and merge the sparse tail
($\geq 6$ steps) into a single ``6+'' bin; the single-step questions are
excluded as their reference annotations are noisy.
Figure~\ref{fig:acc-lat-by-steps} reports answer accuracy and response latency per group for the baseline (\textit{EarlyReasoning [7]}) and the proposed \textit{EarlyReasoning w/
RetroThinking \& DPO}.

Both systems exhibit the expected decay in accuracy as reasoning complexity grows, dropping from the 2-step regime to near-zero performance at 6+ steps.
Across every complexity bin, RetroThinking with DPO improves over the baseline.
The gains are largest and most reliable in the mid-complexity regime.
At 3 steps accuracy rises from $23.1\%$ to $37.9\%$ ($+14.8$ points), and at 4 steps from $12.8\%$ to $25.2\%$ ($+12.4$ points), with non-overlapping error bars in both cases.
Improvements at 5 and 6+ steps are positive but smaller and fall within the margin of error, indicating that the method primarily recovers problems of moderate difficulty rather than the hardest long-horizon questions, where both systems remain weak.

Response latency increases monotonically with the number of reasoning steps for both systems, consistent with longer reasoning chains requiring more generation before an answer is produced.
RetroThinking with DPO incurs a modest latency overhead that widens with complexity, from a negligible difference at 2--3 steps to roughly $+1.2$\,s at 4 steps and $+1.4$\,s at the 6+ bin.
We attribute this to the additional retrospective reasoning the model performs on harder problems, which is precisely where the accuracy gains are
concentrated.


\section{Conclusion}
In this work, we introduced \textit{RetroThinker}, a novel post-training framework that equips streaming SpeechLLMs with forward-only retrospective reasoning capabilities. Our experiments on the spoken GSM8K benchmark demonstrate that models such as Moshi can be effectively fine-tuned to self-verify Chain-of-Thought (CoT) steps and generate corrective continuations without requiring trace rollback.  
Crucially, while retrospective training improves accuracy at the cost of increased latency, we successfully mitigate this overhead by integrating Early Reasoning and LengthDPO.
Furthermore, our comprehensive ablation studies validate the efficacy of our three-stage pipeline and substantiate our core hypotheses. By analyzing current failure modes, we highlight promising avenues for future research, ultimately advancing the development of retrospective, low-latency reasoning in spoken language models.

\section{Limitations}
Our evaluation is limited to a TTS version of GSM8K.
This benchmark is useful because it provides step-level rationales and objective answers, but it does not capture the acoustic variability, interruptions, disfluencies, or task diversity of natural spoken dialogue.
Therefore, our claims should be interpreted as evidence for streaming speech math reasoning rather than broad conversational generalization.

Our accuracy metric also depends on Whisper transcription followed by an LLM judge.
We use this pipeline to evaluate the actual user-facing waveform and to remain comparable with prior work, but ASR or judge errors can still affect absolute accuracy.
Because every system is evaluated with the same pipeline, the relative comparisons are more reliable than the absolute scores.

Finally, while RetroThinker significantly improves performance, it does not entirely eliminate the latency cost of reasoning. Before DPO, retrospective corrections add noticeable delay, and even after DPO, the QC threshold no longer controls latency as precisely as in non-retrospective early reasoning. Furthermore, resolving highly complex, multi-step problems remains an open challenge for our model, reflecting a broader difficulty in streaming speech reasoning. Future work will focus on testing natural and noisy speech benchmarks, as well as developing methods to explicitly control the number and length of retrospective corrections.

\section{AI-Generated Content Disclosure}
Generative AI tools were utilized exclusively for figure formatting, language editing and refinement of this manuscript.
The authors take full responsibility for all scientific content, originality, and the final submitted work.

\section{Acknowledgments}
We thank Wei Zhou for helpful discussions during the early stage of this project.

\bibliographystyle{IEEEtran}
\bibliography{mybib}

\end{document}